\documentstyle[aps,prb,multicol,epsf,graphicx]{revtex}

\newcommand \bea {\begin{eqnarray}}
\newcommand \eea {\end{eqnarray}}
\newcommand \be {\begin{equation}}
\newcommand \ee {\end{equation}}

\newcommand \bi {\bibitem}
\newcommand \s {\sigma}

\begin{document}

\title{Random energy levels and low-temperature expansions for spin glasses} 

\author{M. Picco$^{(1)}$, F. Ritort$^{(1,2)}$ and M. Sales$^{(2)}$} 

\address{(1) LPTHE, Universit\'e Pierre et Marie Curie, Paris VI\\
        et Universit\'e Denis Diderot, Paris VII\\
        Boite 126, Tour 16, 1$^{\it er}$ \'etage, 4 place Jussieu\\
        F-75252 Paris Cedex 05, France\\
        (2) Departament de F\'{\i}sica Fonamental,
        Facultat de F\'{\i}sica, Universitat de Barcelona\\ 
        Diagonal 647, 08028 Barcelona, Spain
}


\maketitle
\begin{abstract}
In a previous paper (cond-mat/0106554) we showed the existence of two
new zero-temperature exponents ($\lambda$ and $\theta'$) in two
dimensional Gaussian spin glasses.  Here we introduce a novel
low-temperature expansion for spin glasses expressed in terms of the
gap probability distributions for successive energy levels. After
presenting the numerical evidence in favor of a random-energy levels
scenario, we analyze the main consequences on the low-temperature
equilibrium behavior. We find that the specific heat is anomalous at
low-temperatures $c\sim T^{\alpha}$ with $\alpha=-d/\theta'$ which
turns out to be linear for the case $\theta'=-d$.
\end{abstract}

\vspace{.2cm}

Spin glasses are random systems where frustration plays a very
important role.  In a previous paper (hereafter referred as I)
\cite{I} we have shown the importance of considering both small and
large-scale excitations to properly understand the low-temperature
behavior of spin glasses. The thermal exponent $\theta$, which
determines the typical free energy cost to overturn a droplet of large
size, scales like $L^{\theta}$ with $\theta=\theta'+d\lambda$. The
exponent $\theta'$ gives the energy cost associated to the lowest
excitation while the other exponent $\lambda$ has an entropic origin
and accounts for the probability (proportional to $1/V^{\lambda}$) to
find a large-scale lowest excitation.

In standard phenomenological approaches (domain-wall theory or the
droplet model \cite{MILLAN1,DROPLET,FH}) the exponent $\lambda$ is not
necessary because only typical excitations are considered for the
low-temperature behavior. This case corresponds to $\lambda=0$ or,
equivalently, $\theta'=\theta$. Therefore, the main difference between
our approach and the domain-wall approach is that the excitations we
consider in our analysis are not typical at finite temperatures while
those generated in domain-wall theory (by measuring the energy change
in the ground state configuration after a twist of the boundary
conditions, see \cite{VARIOS}) are supposed to be typical. In our
approach we infer the statistical properties of the typical
excitations by looking at those excitations in the extreme tail of the
energy gap distribution. The ultimate reason behind the validity of
our approach is the random character of energy levels. Evidence in
favor of a random energy levels scenario was already presented in I.
In I, besides showing how the exact investigation of the lowest
excitations may identify the two exponents, we also showed how the gap
distribution does not depend on the size of the excitation, justifying
a scenario of random-energy levels. Here we want to go further and
check its validity by extending the analysis to second order
excitations.  Furthermore we want to show also how we can derive a
novel low-temperature expansion for spin glasses and infer some
results from that expansion by assuming random-energy levels. This
expansion is useful to understand the low-temperature behavior of some
quantities such as the specific heat or the spin-glass susceptibility.

Although the main results here and in I only considered 2D Gaussian
spin glasses we have founded reasons to believe that the validity of
our assumptions extends also to higher dimensions. To validate this
new scenario beyond two dimensions we need systematic and powerful
algorithms to look for low-lying excitations. Recent numerical
developments promise a fast growth of this area and studies in 3D will
be crucial \cite{OLIVIER}.

The random-energy levels scenario is based on two assumptions. {\em
Assumption A:} correlations between different energy gaps vanish in
the $L\to\infty$ limit and {\em Assumption B:} Correlations between
excitation volumes $v_i$ and gaps $ E_i$ vanish in the $L\to\infty$
limit. These two assumptions are very natural.  In a random system the
ground state configuration is completely disordered. The fact that
there are large-scale higher excitations indicates that there exist
configurations very close in energy to the ground state energy but
very far from each other in phase space. If the ground state
configuration is suppressed from the set of allowed configurations the
statistical properties of the new lowest gap $E_2-E_1$ will remain the
same to those of the original lowest gap $E_1$. Hence, any variable
($v$ or $E$) appears to be uncorrelated to the energy. Nevertheless,
note that while energy levels are random, excitation volumes may be
correlated among themselves. In a disordered system, the presence of
random-energy levels extends to higher energy typical excitations
supporting the core of the results presented in I. Obviously this is
not true in an non-disordered system where the ground state may have a
crystalline structure and typical excitations are not statistically
represented by the lowest ones. The validity of this description
in terms of a random-energy levels scenario is probably related to the
stochastic stability property of disordered systems \cite{SILVIO}. If
the system is stochastically stable then we can perturb it with a
random Hamiltonian without changing the physical properties of the
system (for instance, the value of the thermal exponent $\theta$). A
slight perturbation of the original Hamiltonian corresponds to shift
and mix the original distribution of energy levels. A good selection
of the type of random perturbation and an appropriate tunning of its
intensity might have the effect of making typical those lowest order
excitations which originally were not. Because the physical properties
of the new perturbed system remain unchanged this implies that the
statistical properties of lowest and typical excitations must be also
the same.

\begin{figure}[tbp]
\begin{center}
\rotatebox{270}{
\includegraphics*[width=5.cm,height=8cm]{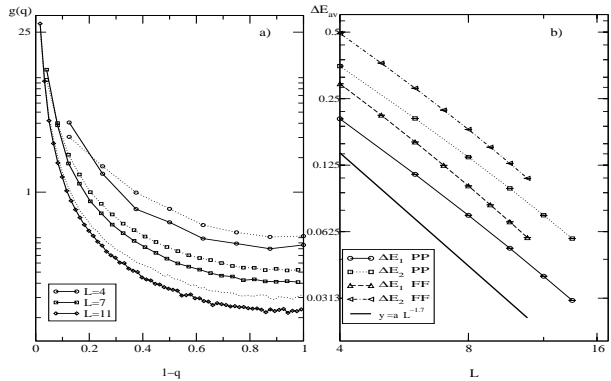}}
\vskip 0.05in
\caption{In plot a) we show the $g(q),g^{(2)}(q)$ for first (solid
line) and second (dashed line) excitations for lattice sizes $4,7$ and
$11$ in the PP case. In plot b) we show, in a log-log scale, the
average gap for the first and second excitations for the PP and FF
cases.The full straight line corresponds to the power law $L^{-1.7}$.}
\label{fig4}
\end{center}
\end{figure}

To verify these assumptions we studied first and second excitations in
the two dimensional Gaussian spin glass defined by

\be
{\cal H}=-\sum_{i<j}J_{ij}\,\s_i\,\s_j~~~~,
\label{eq0}
\ee where the $\s_i$ are the spins ($\pm 1)$ and the $J_{ij}$ are
quenched random variables extracted from a Gaussian distribution with
zero mean and unit variance. In I we found that two exponents
$\lambda$ and $\theta'$ characterize the level spectrum. A question
not addressed in detail was the issue of correlations. Using the
transfer matrix method we have looked at first and second excitations
for 2D spin glasses with periodic-periodic (PP), free-free (FF) and
free-periodic (FP) boundary conditions. Second excitations are also
cluster excitations \cite{FOOTNOTE2} which strongly simplifies the
analysis.  Investigating higher energy levels requires larger
computational effort but can be afforded. Sizes range from $L=4$ up to
$L=11$ for PP and up to $L=14$ for FP and FF. The typical number of
samples is $10^6$ for all sizes. Let $v_1^{(s)}$ and $v_2^{(s)}$
denote the volume of the first and second excitations with respective
gaps $ E_1^{(s)}, E_2^{(s)}$ for a given sample $s$. We define the
second-excitation (with energy gap $ E_2$) probability distribution
$P^{(2)}(v, E_2)=g^{(2)}(v)\,\hat{P}^{(2)}_v( E_2)$ in analogous way
as we did in I for first excitations. In figure~\ref{fig4} we show the
probability to find a second excitation with volume $v$ $g^{(2)}(v)$
for the PP case (the other cases are similar) showing the same
functional shape to that found in the case of the lowest first
excitations. Also in that figure we show the average gap for second
excitations compared to the first-excitation gap as function of $L$
for PP and FF. We point out two important features: One the one hand,
$g^{(2)}(v)$ follows the same functional form to the one describing
first excitations with the same exponent $\lambda$ and tends to $g(v)$
in the large volume limit. It is also remarkable that the
distributions $g(v)$ and $g^{(2)}(v)$ are nearly equal, except from a
small discrepancy in the single-spin excitations weight. The reason
for this difference comes from the excluded volume effect arising from
the single-spin first excitation which leaves a smaller volume $V-1$
available to the second excitation. For small systems this implies a
net decrease of the probability of having one-spin excitations
whereas, larger volume second excitations, are insensitive to this
effect. Furthermore, this effect yields important $v_1-v_2$
correlations. On the other hand, as was the case for for
$\hat{P}_{v_1}( E_1)$, the energy distribution $\hat{P}^{(2)}_{v_2}(
E_2)$ is independent of the size $v_2$ of the second excitation. As
expected we have that $\hat{P}^{(2)}_v( E_2)=L^{-\theta ''}\hat {\cal
P}^{(2)}( E_2/L^{\theta ''})$. Within numerical precision $\theta
''=\theta '$ for a given size.

We now focus on the issue of the existence of correlations. Let us
denote by $x_s,y_s$ any two quantities for sample $s$ and let us
consider their corresponding correlation,

\be
C_{x,y}(L)=\frac{\overline{x_sy_s}-\overline{x_s}\,\,\overline{y_s}} 
{\sqrt{\overline{x_s^2}  
-(\overline{x_s})^2}\sqrt{\overline{y_s^2}
-(\overline{y_s})^2}}~~~~.
\label{eq9}
\ee

\noindent
We find that correlations between energies and volumes of the type
$C_{v^{(i)}, E^{(j)}}(L)$, whichever they are (first or second
excitations, i.e. $i,j=1,2$) asymptotically vanish in the $L\to\infty$
limit (figure 2a). Correlations between the energies of the levels
deserve some comments. Since $ E_2 > E_1$ there are trivial energy
correlations. In this case it is convenient to consider correlations
in (\ref{eq9}) taking $x=\Delta_1= E_1$, $y=\Delta_2= E_2- E_1$. We
see a slow but systematic decrease of $C_{\Delta_1,\Delta_2}(L)$ with
$L$ suggesting there are no gap-gap correlations (figure 2b).
Nevertheless correlations of the type $C_{v_1,v_2}(L)$ are much higher
and, although they saturate and show a tendency to decrease later, we
have no evidence that they indeed vanish in the $L\to\infty$ limit
(figure 2c).

\begin{figure}[tbp]
\begin{center}
\rotatebox{270}
{\includegraphics*[width=6cm,height=10cm]{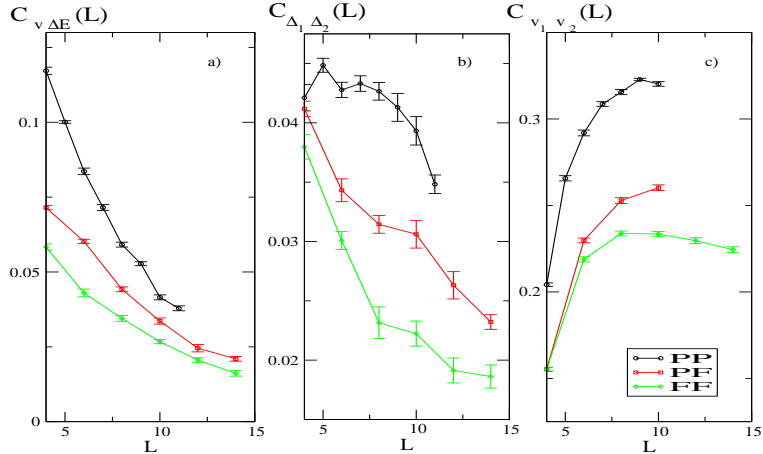}}
\vskip 0.05in
\caption{Different cross-correlations between excitation volumes and energy 
gaps showing that energy levels are random variables.
\label{fig1}}
\end{center}
\end{figure}

Based on the evidence in favor of a random-energy levels scenario we
derive a novel low-temperature expansion for spin glasses. This
expansion has been partially discussed in several papers but, to our
knowledge, has never been worked out in detail. The systematic
procedure to built this expansion is as follows. Consider a spin-glass
system described by a Hamiltonian ${\cal H}_J(\sigma)$ where $\sigma$
denotes the set of discrete spin variables and $J$ are the couplings
that we take continuous in order to avoid accidental degeneracy of the
ground state. Consider a given sample which we will denote by $(s)$
and let us denote its different excited levels by the index
$r=0,1,2,...$ where $r=0$ denotes the ground state configuration,
$r=1$ the first or lowest excitation, $r=2$ the second excitation and
so on.

The low-temperature expansion is done by fixing the ''excited'' level
$r$ and keeping in the partition function all first $r$ levels plus
the ground state configuration. The mathematical object in terms of
which the expansion is written is the probability distribution ${\cal
P}^{(r)}( E_1, E_2,..., E_{r},{\cal C}(v_1,v_2,..,v_r))$ where $ E_i$
is the gap of the $i^{th}$ excitation and ${\cal C}(v_1,v_2,..,v_r)$
stands for a set of variables including all excitation volumes $v_i$
as well as some other volumes obtained by a given number of set
operations (for instance, unions and intersections among the
$v_i$). This set ${\cal C}$ of volume variables may be quite complex
and strongly depend on the observable we are expanding. For $r=1$ the
appropriate probability is ${\cal P}^{(1)}( E_1,v_1)$ as considered in
I. In practice, this procedure generates a low-temperature expansion
in powers of $T$ up to order $T^r$. Including higher-order excitations
in the partition function yields always higher order $T$ corrections
so the expansion is well defined.  As an example, let us see how the
expansion works for the spin-glass susceptibility
$\chi_{SG}=V\overline{\langle q^2\rangle}$ up to order $T^2$ ($q$ is
the overlap between two replicas). In this case $r=2$ and the
probability is ${\cal P}^{(2)}( E_1, E_2,v_1,v_2,v)$ where $v=v_1 \cup
v_2- v_1\cap v_2$, so ${\cal C}(v_1,v_2)$ is the set of variables
including each of the excitation volumes plus their total
non-overlapping volume. This probability can be written as ${\cal
P}^{(2)}( E_1, E_2,v_1,v_2,v)=g_{v_1,v_2,v}\hat{P}^{(2)}_{v_1,v_2,v}(
E_1, E_2)$ this last term being the conditioned probability for a
given triplet $v_1,v_2,v$ to have energy gaps $ E_1, E_2$.  Therefore,
if we keep only the first and second excitations in the Hamiltonian
and we denote $x_i=\exp(-\beta E_i)$, we get

\bea \chi_{SG}=V-\frac{8}{V}\sum_{v_1}
v_1(V-v_1)\int_0^{\infty}d E_1\int_{
E_1}^{\infty}d E_2 \frac{x_1{\cal P}^{(2)}_{v_1}
(E_1, E_2)}{(1+x_1+x_2)^2}-\nonumber\\
 \frac{8}{V}\sum_{v_2}
v_2(V-v_2)\int_0^{\infty}d E_1\int_{
E_1}^{\infty}d E_2 \frac{x_2{\cal P}^{(2)}_{v_2}
(E_1, E_2)}{(1+x_1+x_2)^2}-
\nonumber\\\frac{8}{V}\sum_{v}
v(V-v)\int_0^{\infty}d E_1\int_{
E_1}^{\infty}d E_2 \frac{x_1x_2{\cal P}^{(2)}_{v}
(E_1, E_2)}{(1+x_1+x_2)^2}~~~,
\label{eq1}
\eea

\noindent
where we have defined ${\cal P}^{(2)}_{v_i}( E_1, E_2)=\sum_{{\rm
all~}v's{\rm~except~}v_i}{\cal P}^{(2)}( E_1, E_2,v_1,v_2,v)$ where
$v_i$ stands for $v_1,v_2$ or $v$. Expression (\ref{eq1}) can be
worked out in the limit $\beta\to \infty$ yielding, after some lengthy
calculations,

\bea \chi_{SG} =V-\frac{4T}{V}\sum_{v_1}v_1(V-v_1) {\cal
P}^{(1)}(0,v_1)-\frac{8\log(2)T^2}{V}\sum_{v_1}v_1(V-v_1)
{\cal P}'^{(1)}(0,v_1)+\frac{4T^2}{V}\Bigl(\log(6)\nonumber\\
\sum_{v_1}{\cal P}^{(2)}_{v_1}(0,0)v_1(V-v_1) -2\log(3/2)
\sum_{v_2}{\cal P}^{(2)}_{v_2}(0,0)v_2(V-v_2) -\log(4/3)
\sum_{v}{\cal P}^{(2)}_{v}(0,0)v(V-v)\Bigr)+{\cal O}(T^3)
\label {eq2}
\eea

\noindent
where ${\cal P}^{(1)}( E_1,v_1)=\int_{0}^{\infty}d( E_2){\cal
P}^{(2)}_{v_1}( E_1, E_2),{\cal P}^{(1)}( E_1,v_1)=\frac{\partial
{\cal P'}^{(1)}( E_1,v_1)}{\partial E_1}$, and these expressions
appear evaluated at $ E_1=0$ in (\ref{eq2}). In (\ref{eq2}) we
recognize the linear contribution in $T$ presented in I. At first
glance, this expansion looks too complicated to be useful. But as we
will see below, it can be properly interpreted in a scenario of
random-energy levels.

One of the most striking consequences of the random-energy levels
scenario is that it can be used to predict the specific heat
exponent. There has been a lot of work to understand the specific heat
anomaly in structural glasses showing that, in good approximation,
specific heat is linear in $T$. This is usually explained by the fact
that this kind of systems have a finite density of states at zero gap
\cite{GLASSES}. Contrarily, in spin glasses the behavior of the
specific heat at low-temperatures has not received much attention
probably because the question (both from the numerical and
experimental point of view) can be hardly answered due to the
difficulty to reach thermal equilibrium at low temperatures.  Cheung
and McMillan \cite{CM} claimed that the specific heat in 2D should be
linear in $T$ while Fisher and Huse \cite{FH} made some observations
about the sample to sample fluctuations of the ground state energy and
its relation to the low-temperature specific heat. The specific heat
is given by the usual fluctuation-dissipation formula,
$c=\frac{\beta^2}{V}(\overline{<E^2>-<E>^2})$.  Using the low-$T$
expansion method discussed before we expand $c$ up to any order in
$T$. The calculations are less laborious than for $\chi_{SG}$ because
the relevant probability functions ${\cal P}^{(r)}( E_1, E_2,...,
E_{r})$ do not depend on the spectrum of excitation volumes. If we
expand up to order $T^2$ then we must consider only the two
lowest-lying excitations in the Hamiltonian. This yields the following
expression for the specific heat,

\begin{eqnarray}
c(T,L)=\frac{\beta^2}{V}\int_0^{\infty}dx_1\int_{x_1}^{\infty}dx_2{\cal
P}^{(2)}(x_1,x_2)\Bigl(\frac{x_1^2\exp(-\beta x_1) 
+(x_1-x_2)^2 \exp(-\beta(x_1+x_2))+x_2^2\exp(-\beta
x_2)}{\left(1+\exp(-\beta x_1)+\exp(-\beta x_2)\right)^2}\Bigr).
\label{eq14}
\end{eqnarray}   

In the limit $T\to 0$ after some calculations we obtain,

\be 
c(T,L)=\frac{\pi^2T{\cal P}^{(1)}(0)}{6V}+\frac{9\zeta(3)T^2{\cal
P}'^{(1)}(0)}{2V}+\frac{0.77564\, T^2{\cal P}^{(2)}(0,0)}{V}
\label{eqc}
\ee
where $\zeta(s)=\sum_{k=1}^{\infty}k^{-s}$ is the Riemann function and
${\cal P}'^{(1)}( E)=\frac{\partial{\cal P}^{(1)}( E)}{\partial
E}$. In general, it can be shown that the terms appearing in the
expansion at order $T^r$ are of the type, $\nabla_{ E_1,.., E_v}{\cal
P}^{(u)}( E_1,.., E_u)$ evaluated at $ E_i=0\,,\forall i$ with $u+v=r$
(the symbol $\nabla$ denotes all possible partial derivatives).

According to the assumption A of the random-energy levels scenario the
$r$-point energy probability distribution factorizes, i.e. ${\cal
P}^{(r)}( E_1, E_2,..., E_{r})=\prod_{i=1}^r {\cal P}_i^{(1)}( E_i)$
where all the ${\cal P}_i^{(1)}( E_i)$ scale with the size of the
system $L$ with the same exponent $\theta'$,

\be
{\cal P}_i^{(1)}( E_i)=L^{-\theta'}\hat{{\cal P}}_i\Bigl (\frac{
E_i}{L^{\theta'}}\Bigr)
\label{eqscal}
\ee

\noindent and the dependence of the energy level $i$ enters only
through the scaling function $\hat P$. The presence of the same
exponent $\theta'$ for all levels is also a consequence of the
random-energy levels scenario. Therefore, at order $T^r$ in the
low-$T$ expansion all terms scale like $(TL^{-\theta'})^r/V$ thus
leading (since $\theta'< 0$) to an apparent divergent series. The
whole series can then be resumed in a singular function $\hat{c}(x)$,
$c(T,L)=\frac{1}{V}\hat{c}(TL^{-\theta'})$ like in ordinary critical
phenomena. In the scaling region $T\to 0$ and $TL^{-\theta'}$ finite,
we have a finite heat-capacity $Vc(T,L)$. But if we take first the
limit $L\to\infty$ and afterwords $T\to 0$ the $L$ dependence must
disappear in the specific heat, therefore $\hat{c}(x)\to
x^{-\frac{d}{\theta'}}$ when $x\to\infty$ yielding $c(T\to 0)\sim
T^{\alpha}$ with $\alpha$ the specific heat exponent
$\alpha=-\frac{d}{\theta'}$. For Gaussian spin glasses in general
dimensions there is a general argument (see I) supporting that
$\theta'=-d$, hence $\alpha=1$. Nevertheless, in I we showed how both
exponents $\theta'$ and $\lambda$ suffer from very strong finite-size
corrections yielding effective values for the exponents $\theta'$ and
$\alpha$ in the range of sizes $L\le 14$: $\theta'_{\rm eff}=-1.7\pm
0.1$ and $\alpha_{\rm eff}\simeq 1.18\pm 0.07$. In figure 2 we show
finite-temperature transfer matrix calculations for the 2D Gaussian
spin glass in the same range of sizes $L\le 14$ which nicely conform
to this prediction (we tried the effective exponents $\theta'_{\rm
eff}=-1.6,\alpha_{\rm eff}=1.25$). Only when the system reaches a
temperature such that $TL^{-\theta'}$ is not too small (i.e. for
pretty large sizes) we will obtain the right linear in $T$ dependence.
It is important to note that for finite systems $c$ is always linear
in $T$ for low enough temperatures. Note that because $\theta'< 0$ the
exponent $\alpha$ is always positive. The calculation of the
low-temperature specific heat exponent $\alpha$ provides an indirect
way to determine the gap exponent $\theta'$. Preliminary
investigations for the SK model~\cite{RS} show that the effective gap
exponent $d/\theta'$ is well compatible with 2 giving the well know
result\cite{TAP} $\alpha=2$.

\begin{figure}[tbp]
\begin{center}
{\includegraphics*[width=8cm,height=6cm]{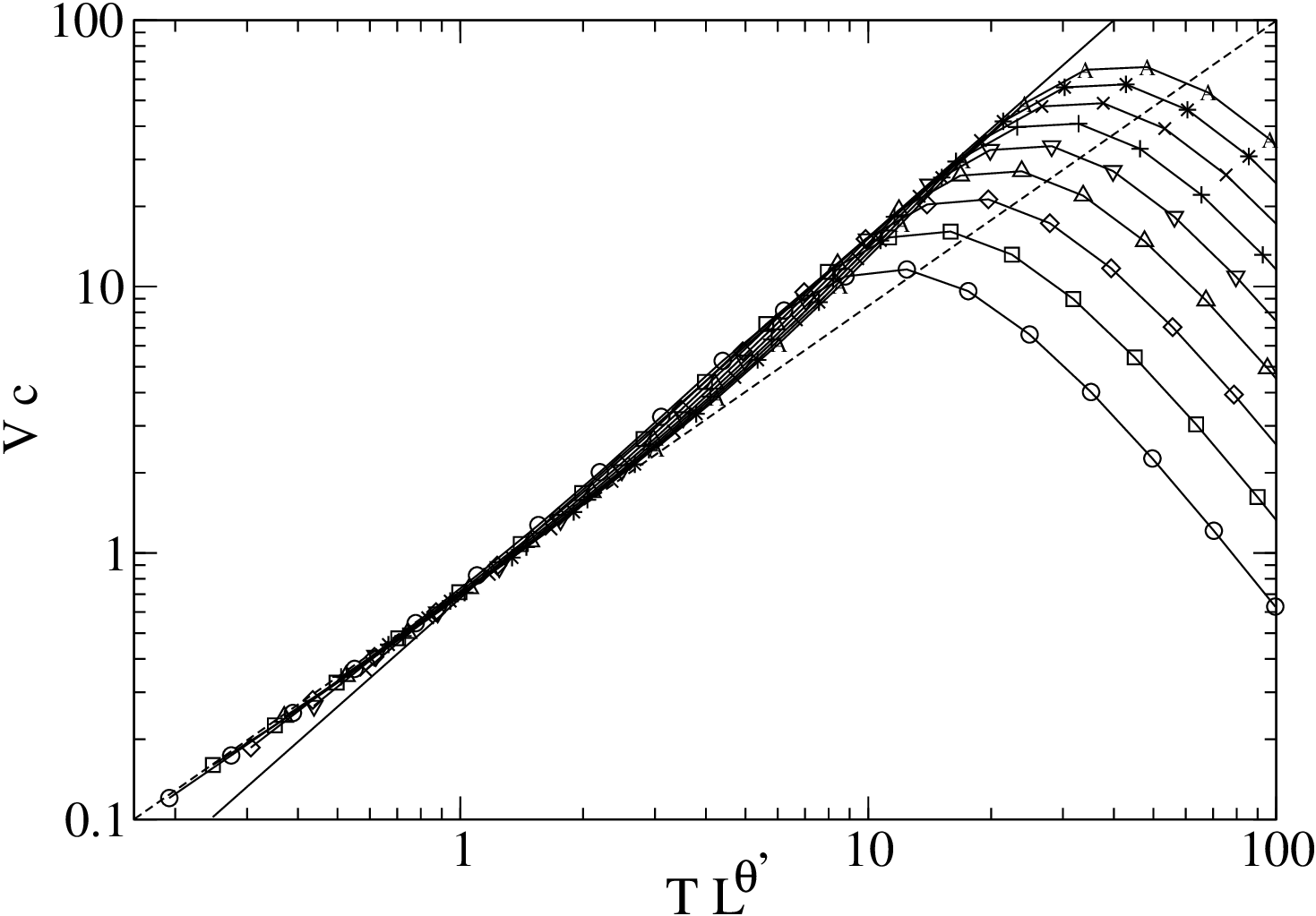}}
\vskip 0.05in
\caption{Heat capacity $Vc(T,L)$ versus $TL^{-\theta'}$
($\theta'=-1.6$) plotted in log-log scale for the FF case with
different lattice sizes $L=6-14$ (bottom to top on the right). We also
show the asymptotic behaviors $\hat{c}(x\gg 1)\sim x^{1.25}$
(full line) and $\hat{c}(x\ll 1)\sim x$ (dashed line).
\label{fig5}}
\end{center}
\end{figure}

We discuss now the behavior of the $\chi_{SG}$ which is certainly more
subtle. As observed in I, the linear term in $T$ scales like
$VTL^{-\theta}$ but the first term of ${\cal O}(T^2)$ in expression
(\ref{eq2}) scales like $VT^2L^{-\theta-\theta'}$. The rest of
quadratic terms would also scale like $VT^2L^{-\theta-\theta'}$ if the
three distributions $g_{v_1},\;g_{v_2},\;g_{v}$ ($v=v_1 \cup v_2-
v_1\cap v_2$) defined by $P_{v_i}( E)=g_{v_i}\hat P_{v_i}( E)$ scale
with the same exponent $\lambda$. Indeed, we have seen that
$g_{v_1},g_{v_2}$ are described by the same exponent $\lambda$ and we
also verified that the same is true for the $g_v$. Therefore, to
understand the character of the low-$T$ expansion it is necessary to
consider higher orders in $T$ for $\chi_{SG}$. The cubic term is more
complex but can be also worked out. In this case, the third excitation
must be included in the calculation and the $T^3$ gets contribution
from the first, second and third excitations. The first and second
excitation will yield terms of the type $VT^3L^{-\theta-2\theta'}$,
but the third excitation, since it can consist of two disconnected
droplets, will yield terms of the type
$VT^3L^{-2\theta-\theta'}$. Therefore, at any order beyond the first
one ($r=1$) there will be terms of the type
$VT^rL^{-u\theta-v\theta'}$ with $u+v=r$, $r>1$; $u,v\ge
0$. Resummation of this divergent series yields,
$\chi_{SG}(T,L)=V\hat{\chi}(TL^{-\theta},TL^{-\theta'})$. There are
two divergent length scales, but since $\theta'\le\theta$
($\theta=\theta'+d\lambda$ with $\lambda\ge 0$) the leading scaling
behavior is governed by the term $TL^{-\theta}$. In renormalization
group language this assertion implies that the leading behavior is
determined by the fixed point which has largest correlation length
exponent.  In the region $T\to 0, L\sim T^{1/\theta}$, the term
$TL^{-\theta'}$ diverges like $L^{\theta-\theta'}$ and the scaling
behavior of the susceptibility is given by,
$\chi_{SG}(T,L)=V\hat{\chi}(TL^{-\theta},\infty)$. For $L\gg
T^{1/\theta}$ and $T\to 0$ this yields the usual low-$T$
result~\cite{JAP} $\chi_{SG}\sim T^{-\gamma}$ with
$\gamma=-\frac{d}{\theta}$ . Sub-leading corrections are then expected
for $\chi_{SG}$ because for a given temperature and finite $L$, even
if we keep $TL^{-\theta}$ finite, the second argument of the scaling
function $\hat{\chi}$ systematically changes with $L$. Only for sizes
large enough such that $TL^{-\theta'}\gg 1$ data would collapse. Note
that these finite-size corrections can be important if $\theta'\ne
\theta$ (i.e. $\lambda>0$) as happens in 2D. This explains the
corrections obtained for the susceptibility exponent $\gamma$ obtained
from Monte Carlo or finite-temperature transfer matrix methods
\cite{CM,HM,JAP}. Actually, in $1D$ (where $\lambda=0$ and
$\theta=\theta'=-1$) it can be shown that the scaling behavior for
small sizes of the spin-glass susceptibility
$\chi_{SG}(T,L)=L\hat{\chi}(TL)$ is nearly perfect.

To summarize, we have shown the numerical evidence in support of
random-energy levels and derived a low-temperature expansion for spin
glasses by progressively including higher excitations into the
partition function. The coefficients in this expansion can then be
written in terms of a set of energy gap probability distributions and
their derivatives evaluated at zero gap.  In particular, we have
obtained the specific heat exponent ($c\sim T^{-\frac{d}{\theta'}}$)
which is linear under the assumption $\theta'=-d$ valid for Gaussian
spin glasses in finite dimensions.  It remains to be seen how 3d spin
glasses fit to the new scenario and how to extend these ideas to $\pm
J$ spin glasses.

{\bf Acknowledgments.} F.R. and M.S. are supported by the Spanish
Ministerio de Ciencia y Tecnolog\'{\i}a, project PB97-0971 and grant
AP98-36523875 respectively. M. P. and F. R. acknowledge support from
the French-Spanish collaboration (Picasso program and Acciones
Integradas HF1998-0097).

\hspace{-2cm}

\end{document}